\documentstyle[11pt,epsfig]{article}
\begin{document}
\title{CMB: The ultimate test for theoretical models aiming at 
describing the very early universe}
\author{\centerline {Mairi Sakellariadou}\\
\\
{\sl Section of Astrophysics, Astronomy, and Mechanics,}\\
{\sl University of Athens, Panepistimiopolis, GR-15784 Zografos, 
Hellas.}\\
\bigskip
}

\maketitle
\begin{abstract}
In this talk, I will illustrate  how one can use the cosmic
microwave background anisotropy measurements, in order to test
theoretical models aiming at describing the early universe.
\end{abstract}

\section{Introduction}
\par 
The cosmic microwave background (CMB) anisotropy provides a powerful
test for theoretical models aiming at describing the early universe.
The characteristics of the CMB anisotropy multipole moments, and more
precisely the position and amplitude of the acoustic peaks, as well as
the statistical properties of the CMB anisotropy, can be used to
discriminate among theoretical models, as well as to constraint the
parameters space. CMB anisotropies are characterized by their angular
power spectrum $C_\ell$, which is the average value of the square of
the coefficients of a spherical harmonic decomposition of the measured 
CMB pattern.
\par
The position and amplitude of the acoustic peaks, as found by the CMB
measurements --- and in particular by the BOOMERanG~\cite{boom},
MAXIMA~\cite{max}, and DASI~\cite{dasi} experiments --- are in
disagreement with the predictions of topological defects models.
More precisely, defects models lead to the following predictions:\\
$\bullet$ Global ${\rm O}(4)$ textures predict the position of the
first acoustic peak at $\ell\simeq 350$ with an amplitude $\sim 1.5$ 
times higher than the Sachs-Wolfe plateau~\cite{rm}.\\
$\bullet$ Global ${\rm O}(N)$ textures in the large $N$ limit
lead to a quite flat spectrum, with a slow decay after $\ell \sim
100$~\cite{dkm}. Similar are the predictions of other global ${\rm O}(N)$ 
defects~\cite{clstrings,num}.\\
$\bullet$ Local cosmic strings predictions are not very well
established and range from an almost flat spectrum~\cite{acdkss} 
to a single wide bump at $\ell \sim 500$~\cite{mark} with extremely 
rapidly decaying tail.\\
In conclusion, the latest CMB anisotropy measurements rule out pure 
topological defects models as the origin of initial density fluctuations.   
\par
As another example, I would like to mention the case of the
pre-big-bang model (PBB), which is a particular cosmological model
inspired by the duality symmetries of string theory. In an isotropic
PBB model with extra dimensions, the amplification of Kalb-Ramnond
axion vacuum fluctuations can lead to large-scale temperature
anisotropies~\cite{dgsv}. Within this model, the perturbations induced
by massless or very light Kalb-Ramond axions lead to a slightly blue
spectrum of isocurvature perturbations~\cite{dgsv}, which in a closed
universe with considerable cosmological constant, can fit the CMB
data~\cite{dmv}.  However, I believe that the existence of the
isocurvature hump~\cite{dmv} can lead to an inconsistency
between the predictions of the PBB model and the CMB measurements, at
least within the current version of this cosmological model. It is not
yet clear whether variations of the PBB model can cure this potential 
inconsistency between theoretical predictions and measurements~\cite{ks}.
\par
The inflationary paradigm is at present the most appealing candidate
for describing the early universe. However, inflation is not free of
open questions and I would like to mention the following three types
of issues, which represent a kind of open questions for any
inflationary model.\\ $\bullet$ It is difficult to implement inflation
in high energy physics. More precisely, the inflaton potential
coupling constant must be very low in order to reproduce the CMB
data. This is related to the question of deciding which kind of
inflationary model is the more natural one.\\ $\bullet$ The quantum
fluctuations are typically generated from sub-Planckian scales and
therefore one should examine the validity of the theoretical
predictions based upon standard quantum mechanics. Recent
studies~\cite{jetal} seem to indicate that inflation is robust to some
changes of the standard laws of physics beyond the Planck scale. \\
$\bullet$ It is almost always assumed that the initial state of the
perturbations is the vacuum. The proof of such a hypothesis, if it
exists, should rely on full quantum gravity, a theory which is still
lacking. If the initial state is not the vacuum, this would imply a
large energy density of inflaton field quanta, not of a cosmological
term type~\cite{ll}. Thus, non-vacuum initial states lead to a
back-reaction problem, which has not been calculated yet.
\par
In this presentation, I would like to briefly comment on the first and
third types of issues, addressed in the context of inflation.
\bigskip
\section{Choice of the inflationary model}
\par 
Since inflation provides, at present, the most appealing candidate for 
describing the early univerese, we face  the choice of the more
physical inflationary scenario, within a rather large variety of different
scenaria.
Following the philosophy that the more natural cosmological model is
the one which arises from particle physics theories, as for example
superstring theories, we will see that topological defects can still
play a role for the measured CMB anisotropies. In addition, within the
context of the CMB anisotropy measurements, one should always keep in
mind the problem of the degeneracy between various cosmological
parameters. In what follows in this section, I will comment on a new
degeneracy apparently arising in the CMB data, that would be due to a
small --- but still significant --- contribution of topological defects.
\par
In many particle physics based models, inflation ends with the
formation of topological defects, and in particular cosmic
strings~\cite{linde2}.  Moreover, cosmic strings are predicted by many
realistic particle physics models.  Thus, even though the current CMB
anisotropy measurements seem to rule out the class of generic
topological defects models as the unique mechanism responsible for the
CMB fluctuations, it is conceivable to consider a mixed perturbation
model, in which the primordial fluctuations are induced by inflation
with a non-negligible topological defects contribution.
\par
We consider~\cite{bprs} a model in which a network of cosmic strings
evolved independently of any pre-existing fluctuation background,
generated by a standard cold dark matter with a non-zero cosmological
constant ($\Lambda$CDM) inflationary phase. As we shall restrict our 
attention to the angular spectrum, we can remain in the linear regime.
Thus,
\begin{equation}
C_\ell =   \alpha     C^{\scriptscriptstyle{\rm I}}_\ell
         + (1-\alpha) C^{\scriptscriptstyle{\rm S}}_\ell~,
\label{cl}
\end{equation}
where $C^{\scriptscriptstyle{\rm I}}_\ell$ and $C^{\scriptscriptstyle
{\rm S}}_\ell$ denote the (COBE normalized) Legendre coefficients due
to adiabatic inflation fluctuations and those stemming from the string
network respectively. The coefficient $\alpha$ in Eq.~(\ref{cl}) is a
free parameter giving the relative amplitude for the two
contributions.  We have to compare the $C_\ell$, given by
Eq.~(\ref{cl}), with data obtained from CMB anisotropy
measurements. In this preliminary work, we do not vary
$C^{\scriptscriptstyle{\rm S}}_\ell$ characteristics and simply use
the model of Ref.~\cite{num}.  
\par
At this point, we should mention that, strictly speaking, the
anisotropy power spectrum reported in Ref.~\cite{num} concerns
theories of global defects. However, we believe that the spectrum of
Ref.~\cite{num} exemplifies the power spectra of both global and
local cosmic strings.  In a future work~\cite{future} we expect to
have the $C_\ell$ induced by cosmic strings, employing a more acurate
numerical simulation of a cosmic strings network.
\begin{figure}
\centering \epsfig{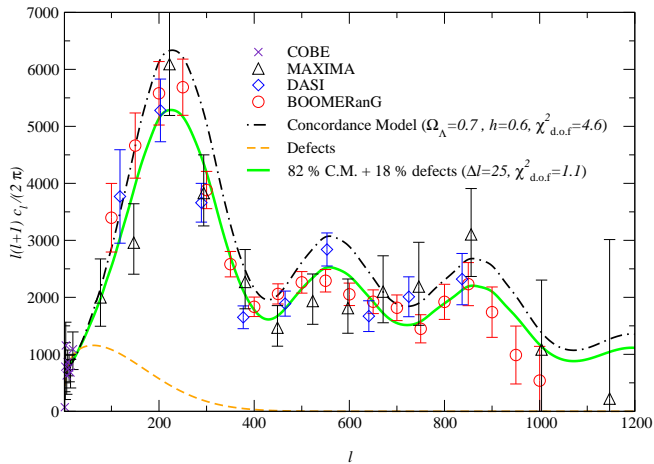}
\caption{$\ell (\ell + 1) C_\ell$ versus $\ell$ for three different
models. The upper dot-dashed line represents the prediction of a
$\Lambda$CDM model, with  $n_{\rm S} = 1$, $\Omega_\Lambda = 0.7$, 
$\Omega_{\rm m} = 0.3$, $\Omega_{\rm b} = 0.05$ and $h = 0.6$.
The lower dashed line is a typical string spectrum. Both lines are
normalized at the COBE scale (crosses).  Combining both curves
with the parameter $\alpha$ produces the solid curve, with a
$\chi^2$ per degree of freedom slightly above unity. The string
contribution is about 18\% of the total.}
\label{fig1}
\end{figure}
Figure~\ref{fig1} shows the two uncorrelated spectra ($\Lambda$CDM
model, strings) as a function of $\ell$, both normalized on the COBE
data, together with the weighted sum. One clearly sees that neither
the upper dot-dashed line --- which represents the prediction of a
$\Lambda$CDM model, with $n_{\rm S} = 1$, $\Omega_\Lambda = 0.7$,
$\Omega_{\rm m} = 0.3$, $\Omega_{\rm b} = 0.05$ and $h = 0.6$ --- 
nor the lower dashed line --- which is a typical string spectrum ---
fit the latest BOOMERanG, MAXIMA and DASI data (circles, triangles 
and diamonds respectively).  The best fit, having $\alpha \sim 0.82$,
yields a non-negligible string contribution, although the 
inflation produced perturbations represent the dominant part for this 
spectrum.
\par
This mixed perturbation model we presented here, leads to the following 
two conclusions~\cite{bprs}:\\
$\bullet$ It seems still a bit premature to rule out any contribution
of cosmic strings to the CMB anisotropy measurements, even though we
are rather confident to conclude that pure topological defects models
are excluded as the mechanism of structure formation.\\
$\bullet$ There is some degree of degeneracy between the model with a
string contribution and the one without strings but with more widely
accepted cosmological parameters. We thus suggest to add the string 
contribution as a new parameter to the standard parameters space.
\bigskip
\section{Non-vacuum initial states for cosmological perturbations of 
quantum mechanical origin} 
\par 
Here, I would like to briefly comment on the choice of the quantum
initial state of the perturbations. In the literature, it is almost
always assumed that the initial state of the perturbations is the
vacuum. The proof of such a hypothesis, if it exists, should rely on
full quantum gravity, a theory which is still lacking. The most
convincing argument in favor of a vacuum initial state is the fact
that non-vacuum initial states imply, in general, a large energy
density of inflaton field quanta, not of a cosmological term
type~\cite{ll}. Thus, one should calculate the back-reaction effect on
the background, but such a computation, even though it is in principle
possible, it has never been performed. Here, instead of relying on
purely theoretical arguments, we will try to
determine~\cite{MRS},~\cite{GMS} whether the vacuum state is the only
state compatible with the observations.
\par
From the theoretical side, the simplest way to generalize the vacuum
initial state, which contains no privileged scale, is to
consider~\cite{MRS} an initial state with a built-in characteristic
scale, $k_{\rm b}$. In a band localized around the prefered scale, the
state contains $n$ quanta, whereas it is still the vacuum elsewhere.
More precisely, we define the state
\begin{equation}
\label{defpsi1}
|\Psi _1(\sigma ,n)\rangle \equiv 
\bigotimes _{{\bf k} \in {\cal D}(\sigma )}|n_{{\bf k}}\rangle 
\bigotimes _{{\bf p} \not\in {\cal D}(\sigma )}|0_{\bf p}\rangle ,
\end{equation}
where the domain ${\cal D}(\sigma )$, in momentum space, is such that,
if ${\bf k}$ is bewteen $0$ and $\sigma $, ${\cal D}$ is filled by $n$
quantas, otherwise ${\cal D}$ is empty. Since the transition between
the emplty and the filled modes is sharp, to ``smooth out'' the state
$|\Psi _1\rangle $, we define a new state $|\Psi _2\rangle $ as a
quantum superposition of $|\Psi _1\rangle $.  In other words,
\begin{equation}
\label{defpsi2}
|\Psi _2(n)\rangle \equiv \int _0^{+\infty }{\rm d}\sigma g(\sigma )
|\Psi _1(\sigma ,n)\rangle ,
\end{equation}
where $g(\sigma )$ is a given function which defines the privileged
scale $k_{\rm b}$. Assuming that the state $|\Psi _2\rangle $ is
normalized, we have $\int _0^{+\infty }g^2(\sigma ){\rm d}\sigma
=1$. 
\par
As it was mentioned earlier, a choice of non-vacuum initial states
suffers from the back-reaction problem~\cite{ll}. Namely, we have the
requirement that the background energy density must be larger than the
energy density due to the quantum fluctuations in the state $|\Psi
_2\rangle $. A back-to-the-envelope estimation~\cite{GMS} indicates
that a model with approximately 60 e-foldings does not suffer from the
back-reaction problem. Thus, there is still a (small) possibility that
non-vacuum initial states are compatible with inflation.
\par
A robust prediction of all models for which the initial state is not
the vacuum is the non-Gaussian character of the induced perturbations.
For models with a prefered scale, the three point (and any
higher-order odd-point) correlation function vanishes, whereas the
four-point (and any higher-order even-point) correlation function does
not satisfy Gaussian statistics~\cite{MRS}.  We will thus
calculate~\cite{GMS} the fourth-order moment (kurtosis) for large
angular scales, for which the source of non-Gaussianity can only be
primordial. Performing a rather lengthy calculation, for a given
ansatz for the function $g(\sigma)$ which defines the previleged scale, 
we obtain~\cite{GMS} the normalised excess of kurtosis ${\cal Q}$,
\begin{equation}
\label{defQ}
{\cal Q}\equiv \biggl\langle\biggl[\frac{\delta T}{T}({\bf e})\biggr]^4
\biggl\rangle/
\biggl\langle\biggl[\frac{\delta T}{T}({\bf e})\biggr]^2\biggr\rangle^2
-3~,
\end{equation}
which we compare to the cosmic variance, calculated for a 
Gaussian field~\cite{cv}, namely
\begin{equation}
{\cal Q}_{\rm CV}\simeq \pm\sigma_{\rm CV}/ 
\biggl\langle\biggl[\frac{\delta T}{T}({\bf e})\biggr]^2\biggr\rangle^2~,
\end{equation}
where
\begin{equation}
\sigma_{\rm CV}^2 = \biggl\langle\biggl[\frac{\delta T}{T}({\bf
e})\biggr]^8 \biggl\rangle - \biggl\langle\biggl[\frac{\delta
T}{T}({\bf e})\biggr]^4 \biggl\rangle^2~.
\end{equation}
The comparison of the excess of kurtosis, calculated from our model,
to the cosmic variance, will indicate the feasibility of detecting
this non-Gaussian signal. Our analysis~\cite{GMS} shows that the
excess of kurtosis is smaller than the cosmic variance. Thus, since we
obtained a signal-to-(teoretical) noise smaller than 1, we conclude
that the non-Gaussian signal, predicted by the choice of non-vacuum
initial states for cosmological perturbations, is not detectable.
\par
The implication of this analysis is quite important for inflation.  We
have examined a possible model, in the context of inflation, which
leads to non-Gaussian statistics of the CMB anisotropy.  We 
showed that this non-Gaussian signature is much smaller than the
cosmic variance, thus it remains undetactable. We therefore 
conclude~\cite{GMS} that the prediction of Gaussian statistics
within the context of inflation is quite robust.
\bigskip
\section{Conclusions}
In this talk, I presented how using the measurements of the CMB
anisotropies, one examines the validity of cosmological models aiming
at describing the early universe. I explained why topological defects
cannot be accepted as the source of primordial fluctuations which gave
rise to structure formation. I also mentioned why the present version
of the pre-big-bang model may soon by in contradiction with the
CMB data. I then showed that a cosmological
model where perturbations are induced by inflation with a
non-negligible topological defects contribution, is compatible with
the latest CMB anisotropy measurements. In addition, as I said, such a
model can lead to a new degeneracy in the data. Finally, I discussed why
the prediction of inflation, namely that the CMB anisotropies obey
Gaussian statistics, is indeed robust.
\bigskip
\section*{Acknowledgments}
It is a pleasure to thank the organizers of the Cosmo 01 meeting for
inviting me to present this work. I would also like to thank all my
colleagues, with whom I collaborated on the projects I presented here.

\end{document}